\documentclass[aps, prx, superscriptaddress, floatfix, twocolumn]{revtex4-2}

\usepackage{amsfonts, amssymb, amsmath}
\usepackage{graphicx}
\usepackage{hyperref}
\usepackage{color,soul}
\usepackage{siunitx}
\usepackage[version=3]{mhchem}
\usepackage{physics}
\usepackage{pdfrender}
\usepackage{mathtools}
\usepackage{cases}
\usepackage{verbatim}
\usepackage{xspace}

\DeclareSIUnit\angstrom{\text {Å}}

\newcommand{\VPASC}{\ensuremath{V_\mathrm{AD}}}
\newcommand{\VKi}{\ensuremath{V_\mathrm{D1}}}
\newcommand{\VKii}{\ensuremath{V_\mathrm{D2}}}
\newcommand{\VKiii}{\ensuremath{V_\mathrm{D3}}}
\newcommand{\PASC}{AD\xspace}
\newcommand{\Ki}{D1\xspace}
\newcommand{\Kii}{D2\xspace}
\newcommand{\Kiii}{D3\xspace}
\newcommand{\VHi}{\ensuremath{V_\mathrm{H1}}}
\newcommand{\VHii}{\ensuremath{V_\mathrm{H2}}}
\newcommand{\VL}{\ensuremath{V_\mathrm{L}}}
\newcommand{\VR}{\ensuremath{V_\mathrm{R}}}
\newcommand{\IL}{\ensuremath{I_\mathrm{L}}}
\newcommand{\IR}{\ensuremath{I_\mathrm{R}}}
\newcommand{\gL}{\ensuremath{g_\mathrm{L}}}
\newcommand{\gR}{\ensuremath{g_\mathrm{R}}}
\newcommand{\EABS}{\ensuremath{E_\mathrm{ABS}}}
\newcommand{\EZ}{\ensuremath{E_\mathrm{Z}}}

\newcommand{\Bx}{\ensuremath{B_x}}
\newcommand{\Bz}{\ensuremath{B_z}}

\begin{document}


\title{Probing Majorana localization of a phase-controlled three-site Kitaev chain\newline with an additional quantum dot}

\author{Alberto~Bordin}
\thanks{These authors contributed equally to this work.}
\author{Florian~J.~Bennebroek~Evertsz'}
\thanks{These authors contributed equally to this work.}
\author{Bart~Roovers}
\thanks{These authors contributed equally to this work.}
\author{Juan~D.~Torres~Luna}
\thanks{These authors contributed equally to this work.}
\author{Wietze~D.~Huisman}
\author{Francesco~Zatelli}
\author{Grzegorz~P.~Mazur}
\author{Sebastiaan~L.~D.~ten~Haaf}
\affiliation{QuTech and Kavli Institute of NanoScience, Delft University of Technology, Delft, The Netherlands}
\author{Ghada~Badawy}
\author{Erik~P.~A.~M.~Bakkers}
\affiliation{Department of Applied Physics, Eindhoven University of Technology, Eindhoven, The Netherlands}
\author{Chun-Xiao~Liu}
\affiliation{QuTech and Kavli Institute of NanoScience, Delft University of Technology, Delft, The Netherlands}
\author{Ruben~Seoane~Souto}
\affiliation{Istituto de Ciencia de Materiales de Madrid (ICMM),  Consejo Superior de Investigaciones Científicas (CSIC), Sor Juana Inés de la Cruz 3, 28049 Madrid, Spain}
\author{Nick~van~Loo}
\author{Leo~P.~Kouwenhoven}
\email{l.p.kouwenhoven@tudelft.nl}
\affiliation{QuTech and Kavli Institute of NanoScience, Delft University of Technology, Delft, The Netherlands}

\date{\today}

\begin{abstract}
Few-site implementations of the Kitaev chain offer a minimal platform to study the emergence and stability of Majorana bound states. Here, we realize two- and three-site chains in semiconducting quantum dots coupled via superconductors, and tune them to the sweet spot where zero-energy Majorana modes appear at the chain ends. We demonstrate control of the superconducting phase through both magnetic field and sweet-spot selection, and fully characterize the excitation spectrum under local and global perturbations. All spectral features are identified using the ideal Kitaev chain model. To assess Majorana localization, we couple the system to an additional quantum dot. The absence of energy splitting at the sweet spot confirms the high quality of the Majorana modes, despite the minimal size of the chains.
\end{abstract}

\maketitle

\section*{Introduction}

\begin{figure*}
    \centering
    \includegraphics[width=\textwidth]{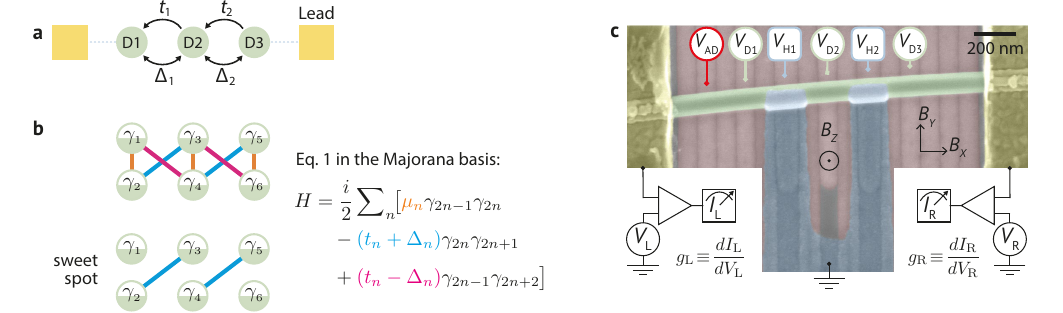}
    \caption{\textbf{Forming Kitaev chains in hybrid nanowires.}  \textbf{a.} Schematic of a three-site Kitaev chain coupled to two metallic leads (yellow). $t_n$ and $\Delta_n$ couple neighbouring quantum dots. \textbf{b.} When all inter-dot couplings are real numbers, a three-site chain can be represented in the Majorana basis with this simple diagram: $\mu_n$ couples the Majoranas within the same site whereas $\left(t_n + \Delta_n \right)$ and $\left(t_n - \Delta_n \right)$ couple Majoranas of neighbouring sites. At the $\mu_n=0$, $t_n=\Delta_n$ sweet spot there is one unpaired Majorana mode at each chain end. \textbf{c.} A false-coloured scanning electron micrograph of the reported device including circuit elements and gate voltage labels. An InSb nanowire (green) is deposited on an array of bottom gates (pink) which can define three quantum dots: D1, D2, and D3. They are coupled by two InSb-Al hybrids~\cite{Bordin2023tunable} which are connected by a superconducting loop (blue), grounded. There is also the possibility of defining an additional quantum dot (AD) on either side. Two Cr/Au contacts (yellow) can probe the local density of states on either side of the device via tunneling spectroscopy.}
    \label{fig:1}
\end{figure*}

The Kitaev chain model is one of the simplest implementations of topology in condensed matter~\cite{Kitaev2001unpaired}. It describes a spinless chain of $N$ fermionic sites ($c_n$) with energies $\mu_n$, nearest-neighbour hoppings $t_n$, and superconducting-like pairings $\Delta_n$~\cite{Sau2012realizing} (Fig.~\ref{fig:1}a,b):
\begin{equation}
    H_N = \sum_{n=1}^N \mu_n c_n^\dagger c_n +\! \sum_{n=1}^{N-1}\! \left( t_n c_n^\dagger c_{n+1} + \Delta_n c_n^\dagger c_{n+1}^\dagger + h.c. \right)
    \label{eq:hamiltonian}
\end{equation}
For $N \rightarrow \infty$, Kitaev showed that the chain hosts two Majorana bound states (MBSs), one at each end of the chain, which are exponentially close to zero energy and topologically protected – meaning that no local perturbation of the Hamiltonian can couple them~\cite{Kitaev2001unpaired}. This makes the Kitaev chain a promising candidate for a robust quantum memory, as the dephasing time would be inversely proportional to the energy splitting of the MBSs~\cite{Boross2022dephasing}.

Remarkably, even minimal Kitaev chains of just two sites can host unpaired MBSs, which are exactly at zero energy if $\abs{t_1}=\abs{\Delta_1}$ and $\mu_1 = \mu_2=0$~\cite{Leijnse2012parity}. However, they are not topologically protected as their energy splits linearly with $\abs{t_1} - \abs{\Delta_1}$, motivating the label of \textit{poor man's} Majoranas~\cite{Leijnse2012parity}.
The transition from the poor man's $(N=2)$ to the topological regime $(N \rightarrow \infty)$ is an active field of research~\cite{Sau2012realizing, Leumer2020excact, Ezawa2024even, Miles2024kitaev, Luethi2025fate, Svensson2024quantum, Dourado2025two, Dourado2025majorana}. The general trend is an increase of the protection from perturbations as the chain is scaled up~\cite{Svensson2024quantum}, but the trajectory is not necessarily monotonic: in particular, three-site chains \textit{could} be worse than two-site ones due to next-nearest-neighbour hoppings~\cite{Miles2024kitaev, Svensson2024quantum, Dourado2025majorana} and four-site chains \textit{could} be worse than three-site ones due to even-odd effects~\cite{Ezawa2024even}.
Recently, two- and three-site Kitaev chains were experimentally realized in hybrid semiconducting–superconducting nanowires~\cite{Dvir2023realization, Bordin2025enhanced} and two-dimensional electron gases~\cite{tenHaaf2024a, tenHaaf2024edge}; attracting substantial experimental~\cite{Bordin2023tunable, Zatelli2024robust, Bordin2024crossed, vanDriel2024charge, Bordin2025impact, vanDriel2024cross} and theoretical~\cite{Liu2022tunable, Tsintzis2022creating, SeoaneSouto2023probing, Liu2023fusion, Liu2024enhancing, Pino2024minimal, Tsintzis2024majorana, Bozkurt2024interaction, Pan2025rabi, Alvarado2024interplay, Liu2025scaling, Benestad2024machine, SeoaneSouto2024majorana, Nitsch2024the} attention to the understanding of the underlying physics.

In this work, we realize two- and three-site chains in a single device, describe their phenomenology, and test the Majorana quality by coupling them to an additional quantum dot. This has both a practical and a fundamental purpose. For technological applications, it is important to understand whether or not three-site Kitaev chains are better than two-site ones and whether it is beneficial to scale up the chain even further. For a fundamental understanding of the onset of topology, it is insightful to investigate how the partial protection from perturbations evolves in these finite-size Kitaev chains. In our previous work~\cite{Bordin2025enhanced}, we investigated the stability against \textit{internal} perturbations of the Hamiltonian terms~\eqref{eq:hamiltonian}. Here, we test the stability against the simplest \textit{external} addition to the Hamiltonian: one extra energy level, provided by an additional quantum dot~\cite{Deng2016majorana, Deng2018nonlocality}. Theory predicts that if the MBSs are well localized at the chain ends, the additional quantum dot can couple only to a single Majorana mode and, therefore, nothing happens to its energy; if, however, there is a finite overlap between the MBS wavefunctions, so that the quantum dot can couple to both, then the MBSs gain a finite energy splitting~\cite{Prada2017measuring, Clarke2017experimentally, SeoaneSouto2023probing}.

\section*{Realization of two- and three-site Kitaev chains}

Our device consists of a semiconducting nanowire (InSb~\cite{Badawy2019high}) deposited on top of an array of bottom gates, separated by a dielectric layer (Fig.~\ref{fig:1}c). Two aluminum strips induce superconductivity in two nanowire sections, forming semiconducting-superconducting hybrids. They are connected in a loop geometry so that an out-of-plane magnetic field $\Bz$ can tune the relative phase difference $\varphi$. Besides the aluminum, which is grounded, there are two additional contacts made of gold, which we use to probe the local density of states via tunneling spectroscopy~\cite{Dvir2023realization}. They are connected to respective voltage sources $\left( \VL,\,\VR \right)$, current meters $\left( \IL,\,\IR \right)$, and standard lockins to measure the differential conductances $\left( \gL \equiv \frac{d\IL}{d\VL}, \, \gR \equiv \frac{d\IR}{d\VR}\right)$. Further nanofabrication details are reported in Supplementary Information.

To form a Kitaev chain, we apply voltages to the bottom gates to define three quantum dots separated by the two hybrid sections~\cite{Bordin2024crossed}. A magnetic field $\Bx = \SI{175}{mT}$ parallel to the nanowire ensures spin polarization of all quantum dots (verified in Fig.~\ref{sup:QD-char-A}). The dot electrochemical potentials $\mu_n$ are controlled by the plunger gate voltages $\VKi$, $\VKii$, and $\VKiii$, respectively, while the inter-dot couplings $t_n$ and $\Delta_n$ are tuned with the hybrid gate voltages $\VHi$ and $\VHii$~\cite{Liu2022tunable, Bordin2023tunable}. There is also the option of forming an additional quantum dot (labelled AD) to test the Majorana localization. This additional quantum dot is used for Fig.~\ref{fig:4}, otherwise there is a single tunneling barrier separating the left gold contact from \Ki. All bottom gate settings are available in the linked repository~\cite{Zenodo2025probing}.

\begin{figure*}
    \centering
    \includegraphics[width=\textwidth]{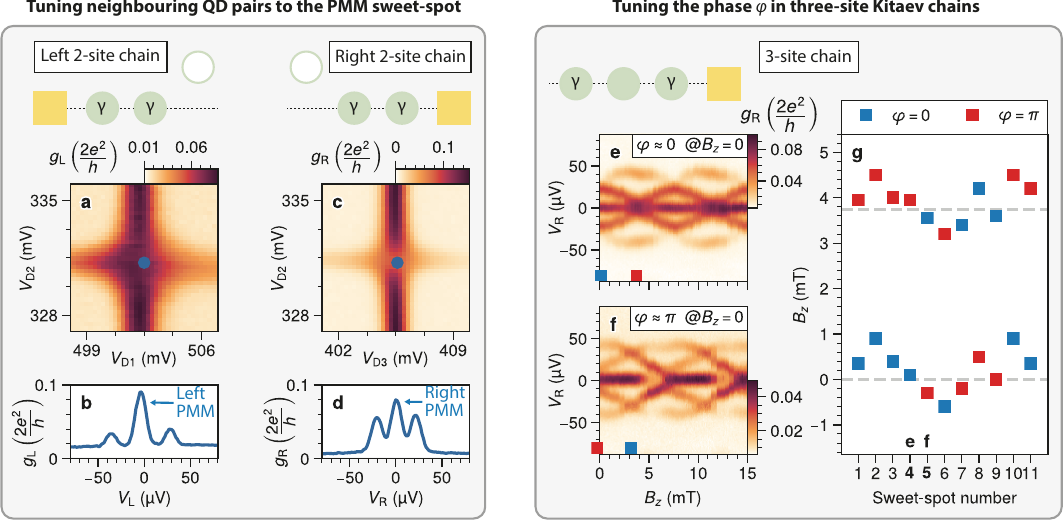}
    \caption{\textbf{Tuning process. a-d.} Two-site chains at the \textit{Poor Man's Majorana} sweet spot. The left chain is measured while keeping \Kiii off-resonance whereas the right chain is measured keeping \Ki off-resonance. Panels \textbf{a} and \textbf{c} show the \Ki-\Kii and \Kii-\Kiii charge stability diagrams measured with standard lockin techniques at zero DC voltage bias ($\VL=\VR=0$). The blue dots mark the points where the finite-bias spectroscopy reported in panels \textbf{b} and \textbf{d} is measured.  \textbf{e-g.} Three-site chains. \textbf{e, f.} Examples of tunneling spectroscopy as a function of the out-of-plane magnetic field $\Bz$. Both panels show a periodic spectrum with $\SI{7.5}{mT}$ period. However, opposite behaviour is seen at $\Bz=0$: in panel~\textbf{e} there is an energy gap while in panel~\textbf{f} the gap is nearly closed. This suggests that at zero out-of-plane field the phase is close to zero in panel~\textbf{e} and close to $\pi$ in panel~\textbf{f}. The zero and $\pi$ points are highlighted with blue and red squares, respectively. The same analysis is repeated multiple times at different Kitaev chain sweet spots (see Fig.~\ref{sup:0-π-inventory} for a full inventory) and summarized in panel~\textbf{g}: overall, at $\Bz=\SI{0}{mT}$ we measured $\varphi\approx0$ seven times and $\varphi\approx\pi$ four times.}
    \label{fig:2}
\end{figure*}

To define two-site Kitaev chains, we use the following procedure. After forming the \Ki, \Kii, and \Kiii dots (their characterization is reported in Fig.~\ref{sup:QD-char-A}), we set \Kiii off-resonance by adding $\sim\!\SI{5}{mV}$ to $\VKiii$. A two-site chain comprising \Ki and \Kii is formed by balancing the $t_1$ and $\Delta_1$ couplings by fine-tuning $\VHi$~\cite{Bordin2023tunable, Dvir2023realization}. The \textit{poor man's Majorana sweet spot} $\abs{t_1} = \abs{\Delta_1}$ is reached when the \Ki-\Kii charge stability diagram of Fig.~\ref{fig:2}a shows a cross shape. The spectrum at the centre of the cross (corresponding to $\mu_1 = \mu_2 = 0$) is reported in Fig.~\ref{fig:2}b and shows a $\abs{2t_1} = \abs{2\Delta_1}\approx\SI{30}{\micro e V}$ energy gap between the zero-bias conductance peak (ZBP) and the first excited state (see Fig.~\ref{sup:fit-gap} for a fit of the spectrum). Similarly, a two-site chain on the right of the device can be formed by setting \Ki off-resonance and balancing the $t_2$ and $\Delta_2$ couplings between \Kii and \Kiii by fine-tuning $\VHii$. At the sweet spot (Fig.~\ref{fig:2}c), we find a $\abs{2t_2} = \abs{2\Delta_2} \approx \SI{20}{\micro e V}$ energy gap (Fig.~\ref{fig:2}d). With these settings, the inter-dot couplings $t_n$ and $\Delta_n$ are much smaller than the Zeeman splitting ($\EZ \gtrsim \SI{200}{\micro V}$, Fig.~\ref{sup:QD-char-A}) and the minimum energy of the Andreev bound states (ABSs) located in the hybrid sections ($\EABS \gtrsim \SI{100}{\micro V}$, Fig.~\ref{sup:PMM-phase-dependence}). 

At this point, it is sufficient to bring \Ki back on resonance to obtain a three-site chain with $\mu_n = 0$ and $\abs{t_n} = \abs{\Delta_n}$ for all $n$. We recall that the $t_n$ and $\Delta_n$ couplings are in principle complex numbers. Their phase is irrelevant in two-site chains, whereas in three-site chains a single non-trivial phase degree of freedom $\varphi$ remains~\cite{Sau2012realizing, Bordin2025enhanced}. This can be tuned with out-of-plane field $\Bz$~\cite{tenHaaf2024edge}, as shown in Fig.~\ref{fig:2}e (the corresponding simulation is reported in Fig.~\ref{sup:20-pack-A}j). We observe a periodic spectrum with a period $T_\varphi=\SI{7.5}{mT}$. At $\Bz=0$ (mod $T_\varphi$) there is a maximum gap separating the ZBP from the first excited state, whereas they merge at $\Bz=T_\varphi/2$ (mod $T_\varphi$). This means that the phase difference is $\approx 0$ at zero flux. This does not seem to be a coincidence: previous works predicted $t_n$ and $\Delta_n$ to be real when there is no magnetic field component perpendicular to the nanowire~\cite{Sau2012realizing, Liu2025scaling}. If all the $t_n$ and $\Delta_n$ are real, then $\varphi$ is either $0$ or $\pi$. Indeed, sometimes we observe $\varphi\approx0$ and sometimes $\varphi\approx\pi$ at $\Bz=0$. For instance, Fig.~\ref{fig:2}f shows an example of $\varphi\approx\pi$ at $\Bz=0$ measured in another three-site sweet spot using different QD and hybrid gate settings. Overall, we characterized the phase of 11 different sweet spots. They are reported in Fig.~\ref{sup:0-π-inventory} and summarized in Fig.~\ref{fig:2}g: at $\Bz=0$, we observed $\varphi\approx0$ seven times and $\varphi\approx\pi$ four times. The standard deviation from zero or $T_\varphi/2$ is $\SI{0.5}{mT}$, which is much smaller than the period. This binary nature of the observed phase behavior has significant implications for scaling to long chains~\cite{Liu2025scaling}. In three-site chains, tuning the phase is straightforward, as a single parameter, 
$\Bz$, suffices. However, in longer chains, each additional site would typically require an extra phase-control parameter $\varphi_n$. Remarkably, Fig.\ref{fig:2}g suggests that the phase difference can be set to $\approx 0$ at $\Bz=0$ by the sweet-spot choice, eliminating the need for additional tuning knobs. This means that an arbitrarily long Kitaev chain can be fully tuned at zero out-of-plane field by selecting, in sequence, sweet spots that preserve $\varphi_n \approx 0$~\cite{Liu2025scaling, Huisman2025}. Importantly, there is no need to tune to zero precisely: it was calculated that, as long as $\varphi_n < \pi/2$ $\forall n$, the Kitaev chain has a finite topological gap~\cite{Liu2025scaling}.

\begin{figure*}
    \centering
    \includegraphics[width=0.96\textwidth]{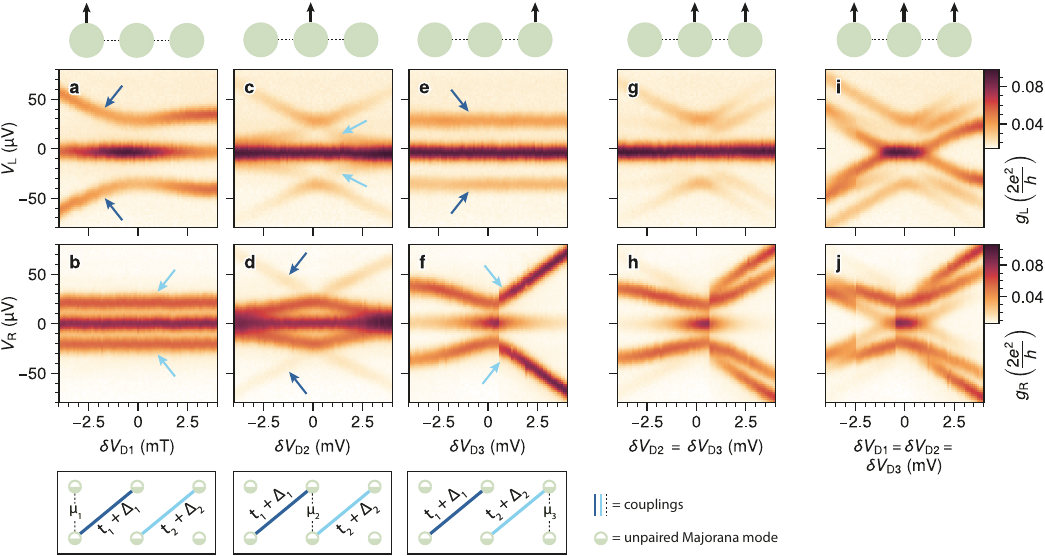}
    \caption{\textbf{Characterization of three-site chain spectra as a function of various QD detunings. a-f.} Conductance spectroscopy of the three-site chain, measured from the left and the right, as a function of each QD plunger gate. The schematics below illustrate the corresponding Kitaev chain model in the Majorana basis. Colour-coded arrows indicate the excited states illustrated in the schematics. \textbf{g, h.} Left and right spectra as \Kii and \Kiii are detuned simultaneously. \textbf{i, j.} Left and right spectra while detuning all QDs of the chain. }
    \label{fig:3}
\end{figure*}

In the rest of the manuscript, we focus on the sweet spot of Fig.~\ref{fig:2}e (number 4 in Fig.~\ref{fig:2}g) and set $\Bz=0$. $\Bx$ is still at $\SI{175}{mT}$ to ensure that all QDs are polarized. The spectra as a function of various QD detunings is reported in Fig.~\ref{fig:3}. The first row shows tunneling spectroscopy from the left lead while the second row shows spectroscopy from the right lead. Different columns show different types of QD detunings: in the first three columns each dot is detuned separately, in the fourth two dots are detuned simultaneously, and in the fifth column all dots are detuned together. As observed in our previous work~\cite{Bordin2025enhanced}, the ZBP persists for any local perturbation of one or even two QDs. Only the global perturbation of three QDs altogether is able to split the ground state degeneracy (Fig.~\ref{fig:2}i,j).

Turning our attention to the excited states, we highlight a feature that might go unnoticed: in the first and the third columns (Fig.~\ref{fig:2}a,b,e,f) there is a single excited state per panel. This reflects the behaviour of an ideal three-site Kitaev model, illustrated in the schematics below. Every site is represented by two Majoranas that couple locally with amplitude $\mu_n$, while $t_n+\Delta_n$ couple neighbouring sites and lead to the excited states marked in dark and light blue~\cite{tenHaaf2024edge}. As long as $\mu_2 = 0$, the excited states are disconnected: a local probe on the leftmost site cannot sense the $t_2+\Delta_2$ state (light blue), whereas a local probe on the rightmost site cannot sense the $t_1+\Delta_1$ state (dark blue). This separation of the excited states breaks down as soon as $\mu_2 \neq 0$ and, indeed, we observe the appearance of a second excited state in each panel~where $\VKii$ is detuned (c, d, g, h, i, and j). In all panels, the second excited state disappears as $\delta \VKii$ approaches $\SI{0}{mV}$.
We note that this disappearance of the second excited state can happen only at $\varphi = 0$ (mod $2\pi$). 
All spectra are reproduced by the numerical conductance simulations with $\varphi=0$ reported in Fig.~\ref{sup:20-pack-A}a-h.

Here, having the $t_1+\Delta_1$ state visible only from the left and the $t_2+\Delta_2$ state visible only from the right is direct evidence of the localization of the excited states. Unfortunately, it does not prove the ground state localization as well. 
Therefore, to investigate the localization of Majorana modes, we rely on another technique: probing the chain with an additional quantum dot~\cite{Prada2017measuring, Clarke2017experimentally, SeoaneSouto2023probing}. This can lift the ground-state degeneracy when coupled to two MBSs, while the system remains degenerate if it couples to one Majorana only.

\section*{Assessing Majorana localization}

\begin{figure*}
    \centering
    \includegraphics[width=0.96\textwidth]{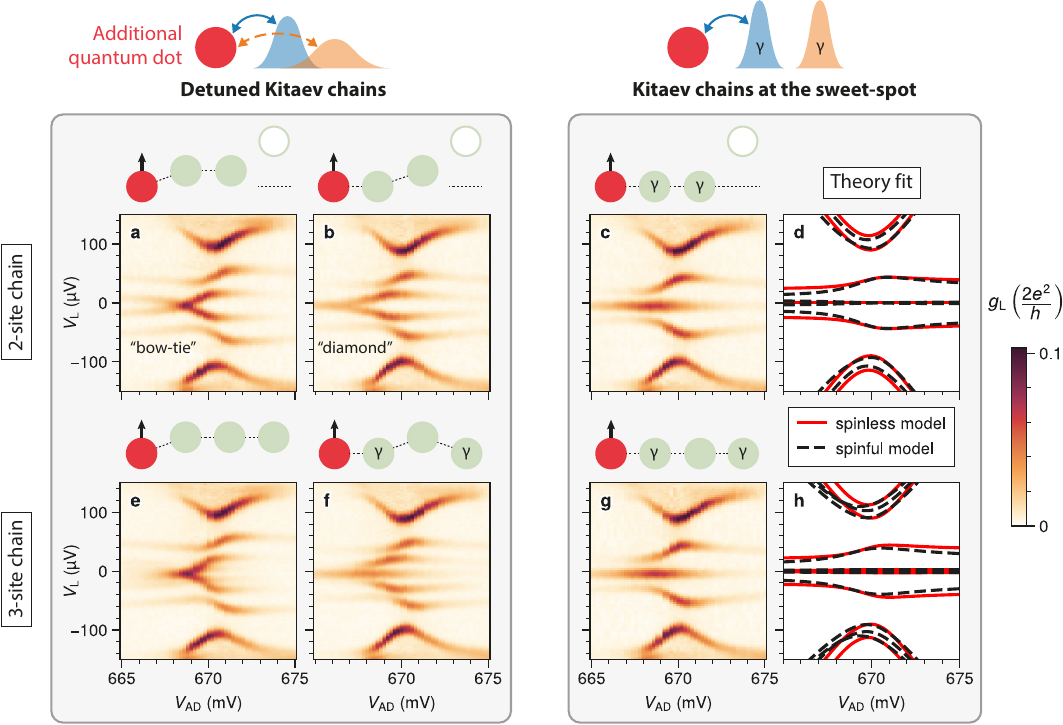}
    \caption{\textbf{Assessing Majorana localization in two- and three-site chains (first and second row, respectively).} Here, the Kitaev chains are coupled to an additional quantum dot, shown in red in the schematics. On the right side, the chains are tuned to the sweet spot. On the left side, they are detuned on purpose in various combinations of $\mu_n\neq 0$.  This simple QD test – measuring the spectrum as a function of the gate voltage $\VPASC$ controlling the additional dot – can expose delocalized Majoranas: if there is a finite Majorana overlap, the coupling to the additional QD splits the ZBP (panels \textbf{a}, \textbf{b}, and \textbf{e}), otherwise, highly localized Majoranas result in persistent ZBPs over the full $\VPASC$ range (panels \textbf{c}, \textbf{f}, \textbf{g}). Panels \textbf{d} and \textbf{h} report a fit of panel~\textbf{c} and \textbf{g} data, respectively, with the spinless model of Eq.~\eqref{eq:hamiltonian} as well as a spinful model discussed in Supplementary Information.}
    \label{fig:4}
\end{figure*}

We form an additional quantum dot on the left of the device (shown in red in the schematics of Fig.~\ref{fig:4}). Its energy levels are controlled by $\VPASC$.
When the QD levels are far from zero energy, the QD acts as a tunneling spectroscopy barrier~\cite{Deng2016majorana}; instead, when a level is close to zero-energy it can couple to the Majoranas localized in the first dot of the chain, \Ki.
Therefore, a simple QD-test involves sweeping $\VPASC$ so that one QD energy level is brought on- and off-resonance. If the unpaired Majorana modes $\gamma_1$ and $\gamma_{2N}$ are perfectly localized at the chain ends, then the ZBP should not be perturbed. If, instead, there is a finite Majorana overlap in \Ki then the Majoranas can couple through the additional dot and, therefore, the ZBP should broaden or even split.

We first look at the scenario where the chains are detuned on purpose (left side of Fig.~\ref{fig:4}), to make sure that we are able to resolve a ZBP splitting. Starting from the two-site chain, we observe in Fig.~\ref{fig:4}a,b a clear splitting of the lowest energy state. As predicted in ref.~\cite{SeoaneSouto2023probing}, this comes in two flavours known as ``bow-tie'' and ``diamond''~\cite{Prada2017measuring}. If both \Ki and \Kii are detuned (Fig.~\ref{fig:4}a), then the ZBP is already split when AD is off-resonance and there is a zero-energy crossing as AD passes through a resonance (bow-tie shape). If only \Kii is detuned (Fig.~\ref{fig:4}b), there is no splitting when AD is off-resonance and a finite splitting when AD is on-resonance (diamond shape). The diamond case is particularly insightful, as the ZBP wouldn't split without the additional quantum dot~\cite{Leijnse2012parity}: if a single QD is detuned from a poor man's Majorana sweet spot, the ZBP revealed by a standard tunneling spectroscopy persists even though the MBSs overlap in one dot~\cite{Dvir2023realization, Zatelli2024robust}. This stresses the strength of this QD-test: it can reveal a local Majorana overlap even where standard tunneling spectroscopy fails to detect it.

On the other hand, the right side of Fig.~\ref{fig:4} shows the situation where our Kitaev chain is tuned to the sweet spot: here, the QD-test does not resolve any ZBP splitting within its linewidth, indicating a strong localization of the Majorana modes.
Theoretical simulations at the sweet spot replicate the spectral dependence.
Furthermore, they can be used to extract microscopic parameters such as the coupling between \PASC and \Ki.
As discussed in Supplementary Information, we fit the spinless model of Eq.~\eqref{eq:hamiltonian} as well as a larger, spinful, model including a second spin species at higher energies~\cite{Tsintzis2024majorana, Luethi2024from}.
Both models qualitatively reproduce the $\VPASC$ dependence. The values of the fitted parameters are reported in Tables~\ref{tab:spinless_fit}  and~\ref{tab:spinful_fit}.
In particular, we extract a strong coupling between \PASC and \Ki, with tunnel amplitudes of order~$\sim \SI{50}{\micro eV}$ and a non-local charging energy $U_\mathrm{nl} \sim \SI{20}{\micro eV}$.

Finally, we turn our attention to the three-site chain case (Figs.~\ref{fig:4}e-h). With a global detuning of all QDs (Fig.~\ref{fig:4}e) we retrieve ZBP splitting similar to the two-site case. However, if only one QD is detuned (Fig.~\ref{fig:4}f), there is no splitting anymore, instead, a ZBP persists over the full $\VPASC$ range. We note that this is not a special property of \Kii: a ZBP persists if \textit{any} of the three-site chain dots is detuned. The \Ki detuning case is trivial (detuning \Ki moves the left Majorana $\gamma_1$ to \Kii, so AD cannot couple to any MBS), the \Kii detuning is shown in Fig.~\ref{fig:4}f, and the \Kiii in Fig.~\ref{fig:4}c. In this sense, the three-site chain is more resilient than two-site chains: Majoranas with a high degree of localization persist even if one of the QDs is off-resonance.

When all QDs are on resonance (Fig.~\ref{fig:4}g) the spectrum looks very similar to the two-site case. In particular, it seems that next-nearest-neighbour couplings are not able to split the ZBP in Fig.~\ref{fig:4}g, as far as this QD-test can resolve, even when looking carefully at the variations of the ZBP linewidth as $\VPASC$ is varied (Fig.~\ref{sup:HWHM}). We do not observe any ZBP broadening in any of the three-site chain sweet spots we tested: the ZBP half-width at half-maximum is $7\!\pm\!\SI{1}{\micro V}$, constantly (Fig~\ref{sup:HWHM}).

All the QD-test phenomenology shown here in Fig.~\ref{fig:4} is reproduced in another Kitaev chain sweet spot within the same device, having different quantum dot orbitals and hybrid gate settings. The corresponding characterization and QD-tests are reported in Supplementary Information (Figs.~\ref{sup:QD-char-B} to~\ref{sup:PC-spin1}).

\section*{Discussion}

\subsection*{Comparison to long nanowires}

A popular strategy for Majorana research involves continuous hybrid nanowires~\cite{Lutchyn2010majorana, Oreg2010helical} rather than QD-based Kitaev chains.
However, the material disorder in long nanowires complicates the unambiguous identification of MBSs in such systems~\cite{Pan2020physical, Prada2020from, Ahn2021estimating, Hess2023trivial}. In contrast, the site-by-site tunability of QD-based Kitaev chains can compensate for material inhomogeneities~\cite{Kouwenhoven2024perspective, Bordin2025engineering}. 
This leads to discrete and localized excitation spectra (Fig.~\ref{fig:3}), where every state can be interpreted with simple models \cite{Kitaev2001unpaired, Sau2012realizing, Tsintzis2024majorana, Dourado2025majorana}. 

With this tunability, we can even simulate disorder – in the form of deliberate perturbations to the system – and study its impact on the spectrum (Figs.~\ref{fig:3} and~\ref{fig:4}).
In short chains, such perturbations can split the ground-state degeneracy. 
However, if the system is perfectly tuned to the sweet spot, the Majoranas are localized in the outer QDs, thus, their wavefunctions do not overlap~\cite{Dourado2025majorana}. This is a fundamental difference compared to continuous nanowires, where the Majorana wavefunctions decay over a characteristic length scale $\xi$~\cite{Kitaev2001unpaired}. This could lead to a detrimental Majorana overlap even in µm-long nanowires~\cite{Stanescu2013dimensional}.

\subsection*{Limitations of our device}

In Figs.~\ref{fig:2} and \ref{fig:3}, we note that the energy gap isn't as large as in other devices~\cite{Zatelli2024robust}, especially on the right side ($\abs{2t_2} = \abs{2\Delta_2} \approx \SI{20}{\micro eV}$). This is limiting the impact of possible next-nearest-neighbour couplings, which scale as $\sim\!\mathcal{O}(t_n/\EZ)$~\cite{Bordin2025enhanced, TorresLuna2025}, but it makes the system more vulnerable to thermal excitations $\sim\!e^{-t_n/k_\mathrm{B} T}$. Finding the best compromise between suppressing next-nearest-neighbour couplings and avoiding thermal excitations is an open question~\cite{TorresLuna2025}.

Regarding the phase dependence, we note finite deviations from 0 or $\pi$ phase at $\Bz = 0$ with a standard deviation $\approx 0.5\,\text{mT}/T_\varphi \approx 0.13\pi$. This does not affect three-site chains, but in long chains it can lead up to a $\approx 50\%$ reduction of the topological gap~\cite{Liu2025scaling}. This can be prevented by stricter sweet-spot selection to have $\varphi_n$ even closer to zero or by better understanding the causes of the deviations. For instance, they could be due to $\Bx$ being not perfectly aligned or the system being not perfectly one-dimensional.
We also note anomalous phase dependences of two-site chains (see Fig.~\ref{sup:PMM-phase-dependence}).

Regarding the QD-test, we note the lack of a direct measurement of the AD spin polarization. In Figs.~\ref{sup:HWHM}, \ref{sup:PC-spin2}, and~\ref{sup:PC-spin1}, the spin is inferred from the theory model. 
Finally, we note the resolution limit of the QD-test, given by tunneling spectroscopy techniques in state-of-the-art dilution refrigerators. Considering the \textit{variations} of the ZBP width as an estimate of the splitting (Fig.~\ref{sup:HWHM}), our resolution is limited to $\sim\!\SI{1}{\micro V}$. Possible ways to go beyond such a resolution include the implementation of cQED spectroscopy techniques or, directly, the creation of a parity qubit made of two coupled Kitaev chains~\cite{Leijnse2012parity, Tsintzis2024majorana, Pan2025rabi}.

\subsection*{Future directions}

In this manuscript, we showed examples of $\varphi\approx 0$ and $\varphi\approx \pi$ at $\Bz = 0$, but we haven’t yet attempted to switch this behaviour deterministically~\cite{Liu2025scaling}. For instance, this could be implemented by switching the QD spins~\cite{Huisman2025}.

Regarding the QD-test, it would be illuminating to study the energy splitting as a function of the Zeeman energy. Next-nearest-neighbour hoppings should become relevant in the $\EZ \sim t_n, \Delta_n$ regime.

On the theory side, we limited the discussion to the qualitative reproduction of the spectra, well captured by the spinless Kitaev model of Eq.~\eqref{eq:hamiltonian}, and the extraction of the Hamiltonian parameters reported in Tables~\ref{tab:spinless_fit} and~\ref{tab:spinful_fit}. In future studies, it can be insightful to quantify the Majorana quality in terms of the Majorana polarization~\cite{Tsintzis2022creating} or other quality measures~\cite{Svensson2024quantum}. This is particularly important with respect to braiding proposals, which have strict Majorana polarization requirements~\cite{Tsintzis2024majorana}.

\section*{Conclusion}

In summary, we realized two- and three-site Kitaev chains within the same nanowire device.  The relative phase $\varphi$ is tuned by out-of-plane magnetic field $\Bz$. 
We also show that it is possible to have either $\varphi\approx 0$ or $\varphi\approx \pi$  at $\Bz = 0$. This demonstrates the possibility of controlling the phase of arbitrarily long Kitaev chains with appropriate sweet-spot choices, eliminating the need of cumbersome flux control~\cite{Liu2025scaling}. 
At $\varphi=0$, we characterized the spectra of three-site chains under local and global perturbation of the quantum dots, finding unprecedented correspondence between the experiment and the Kitaev model, for the ground state and all the excited states.

Finally, we investigated the robustness of Majorana bound states formed at the chain ends against the simplest external perturbation of the Kitaev Hamiltonian: one extra energy level provided by an additional quantum dot.
This QD-test is sensitive to the overlap of the two unpaired Majoranas on one side of the device, even when standard tunneling spectroscopy fails to detect it. Potential overlap of Majorana wavefunctions can cause the ZBP to split, which we do not observe unless the quantum dots of our Kitaev chains are deliberately detuned.

\bigskip
\section*{Acknowledgements}
This work has been supported by the Dutch Organization for Scientific Research (NWO) and Microsoft Corporation Station Q. We acknowledge Michael Wimmer for fruitful discussions and Sasa Gazibegovic for contributions to the nanowire growth. 

\section*{Author contributions}

BR and NvL fabricated the device with help from AB and FZ. FJBE performed the transport measurements with help from AB, FZ, BR, and NvL. JDTL performed the numerical simulations with help from RSS and CXL. GB and EPAMB provided the nanowires. WDH and SLDtH helped understand the phase behaviour. AB, GPM, NvL, and RSS initiated the project. LPK supervised the project. AB and FJBE prepared the manuscript with help from BR, JDTL, NvL, LPK and inputs from all authors. 

\section*{Data availability}
All raw data in the publication, alongside all transport data ever measured on the reported device, and the analysis code used to generate figures are available at~\cite{Zenodo2025probing} \url{https://doi.org/10.5281/zenodo.15168551}.

\bibliography{bib.bib}

\clearpage
\renewcommand\thefigure{S\arabic{figure}}
\renewcommand{\theHfigure}{S\arabic{figure}}
\setcounter{figure}{0}

\onecolumngrid

\begin{center}
  \textbf{\LARGE Supplementary Information}
\end{center}

\bigskip
\section*{Nanofabrication}

The device presented in this work is fabricated using the shadow-wall lithography technique~\cite{Heedt2021shadow, Borsoi2021single}. The nanowire was placed onto a prepatterned substrate, with Ti/Pd gates covered with an ALD-grown dielectric (20 nm Al$_2$O$_3$ and 10 nm HfO$_2$). A 17.5 nm layer of Al was deposited on the nanowire at a 30-degree angle with respect to the substrate, followed by a controlled in-situ oxidation at an O$_2$ overpressure of $\SI{210}{mTorr}$ for 5 minutes. A more detailed description of the Al deposition can be found in Ref.~\cite{Mazur2022spin}. Transene D was used to selectively remove the Al layer outside the nanowire region, which was protected by a PMM layer during the etching. Finally, the Cr/Au ohmic contacts were deposited following Ar milling.

\bigskip
\section*{Theoretical model}
\subsection*{Spinless model}
We extend the spinless Kitaev chain model to include an additional quantum dot as follows:

\begin{align}
    H &= H_N + H_\text{AD} + H_\text{tunnel},\label{eq:spinles}\\
    H_\text{AD} &= (\mu_0 - \mu_\text{offset}) c_0^\dagger c_0,\\
    H_\text{tunnel} &= t_d c_0^\dagger c_1 + \Delta_d c_0^\dagger c_1^\dagger + U_\text{nl} n_0 n_1 + \text{h.c.},
\end{align}
where $H_N$ is given by Eq.~\eqref{eq:hamiltonian}.
The additional quantum dot is described by a single level $c_0$ with chemical potential $\mu_0$, which is connected to $\VPASC$ via the lever arm $\alpha_0$: $(\mu_0 - \mu_\text{offset}) = -e \alpha_0 \delta \VPASC$. In the experimental data, the $\VPASC$ range is not perfectly centred around the \PASC resonance, therefore, we allow for a small chemical potential shift $\mu_\text{offset}$ in the model.
The coupling between \PASC and \Ki is mediated by a normal tunneling term $t_d$ and a phenomenological superconducting term $\Delta_d$. 
Finally, we consider a non-local Coulomb interaction $U_\mathrm{nl}$ between dots \PASC and \Ki. $n_i = c^\dagger_i c_i$ is the number operator at site $i$.

\subsection*{Spinful model}

Complementing the spinless model, where the only source of imperfection for the Majoranas is detuning from the sweet spot, we study the spinful model, that considers additional sources of imperfection~\cite{Tsintzis2022creating}.
In particular, the spinful model includes three additional variables: the Zeeman energy $E_Z$, a local charging energy $U$ on every site, and the spin-orbit angle $\theta$.
For simplicity, we fix the Zeeman and local charging energies to $\EZ = \SI{200}{\micro eV}$ and $U = \SI{3}{meV}$. 
The Hamiltonian for the spinful Kitaev chain coupled to an additional quantum dot is:
\begin{equation}
    H = H_\text{QDs} + H_\text{AD} + H_\text{tunnel},
\end{equation}
where the elements of the Hamiltonian are given by:
\begin{align}
    H_\text{QDs} &= \sum_{i=1}^N \left[ \left(\mu_i + E_Z \right)n_{i\uparrow} + (\mu_i - E_Z) n_{i\downarrow}\right] + U n_{i\uparrow}n_{i\downarrow} + t_i^\text{sc} \left[c_{i\uparrow}^\dagger c_{i+1\uparrow} + c_{i\downarrow}^\dagger c_{i+1\downarrow} \right] + t_i^\text{so} \left[c_{i\uparrow}^\dagger c_{i+1\downarrow} - c_{i\downarrow}^\dagger c_{i+1\uparrow} \right]\notag\\
    &+\Delta_i^\text{sc} \left[c_{i\uparrow}^\dagger c^\dagger_{i+1\uparrow} + c_{i\downarrow}^\dagger c^\dagger_{i+1\downarrow} \right] + \Delta_i^\text{so} \left[c^\dagger_{i\uparrow} c^\dagger_{i+1\downarrow} - c_{i\downarrow}^\dagger c_{i+1\uparrow}^\dagger \right] + \text{h.c.},\\
    H_\text{AD} &=  \left(\mu_0 + E_Z - \mu_\text{offset} \right) n_{0\uparrow} + (\mu_0 - E_Z - \mu_\text{offset}) n_{0\downarrow} + U n_{0\uparrow}n_{0\downarrow},\\
    H_\text{tunnel} &= t_d^\text{sc} \left[c_{0\uparrow}^\dagger c_{1\uparrow} + c_{0\downarrow}^\dagger c_{1\downarrow} \right] + t_d^\text{so} \left[c_{0\uparrow}^\dagger c_{1\downarrow} - c_{0\downarrow}^\dagger c_{1\uparrow} \right]+\Delta_d^\text{sc} \left[c_{0\uparrow}^\dagger c^\dagger_{1\uparrow} + c_{0\downarrow}^\dagger c^\dagger_{1\downarrow} \right] + \Delta_d^\text{so} \left[c^\dagger_{0\uparrow} c^\dagger_{1\downarrow} - c_{0\downarrow}^\dagger c_{1\uparrow}^\dagger \right]\\
    &+ U_\text{nl}(n_{0\uparrow}n_{1\uparrow} + n_{0\downarrow}n_{1\downarrow}+n_{0\uparrow}n_{1\downarrow} + n_{0\downarrow}n_{1\uparrow}) + \text{h.c.}
\end{align}

Inside the chain, the spin-independent hopping  is $t_i^\text{sc} = \tau_i \cos(\theta/2)$ and the spin-orbit hopping is $t_i^\text{so} = \tau_i\sin(\theta/2)$.
Similarly, there are two types of superconducting pairings $\Delta_i^\text{sc} = \eta_i \sin(\theta/2)$ and $\Delta_i^\text{so} = \eta_i \cos(\theta/2)$.
Since we do not know the spin-orbit angle $\theta$, we numerically find the sweet spot for $\mu_i^*(\theta)$ and $\eta_i^*(\theta)$ for a fixed $E_Z$ and $U$ where $0<\theta<\pi$ for $i>0$.
We remark that this model is highly simplified to restrict the number of fitting parameters in the model. For instance, it does not explicitly include the states in the hybrid section, and it assumes uniform $U$ and $\theta$ along the chain. More complete models are worth exploring in future studies. 

\subsection*{Fit of the experimental data}
\begin{figure}[h!]
    \centering
    \includegraphics[width=0.8\textwidth]{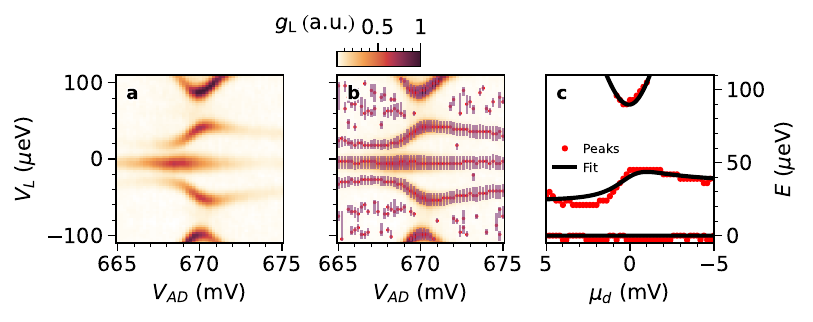}
    \caption{\textbf{Illustration of the fitting procedure of the experimental data. a.} Normalized conductance spectrum from Fig. \ref{fig:4}c. \textbf{b.} Extraction of the peak positions and width from the experimental data in panel \textbf{a}. \textbf{c.} Peaks yelding continuous lines (red dots) and the corresponding fit using Eq.\eqref{eq:spinles} (black lines). }
    \label{sup:fit-method}
\end{figure}
In order to systematically fit the experimental data presented in the main text, we use the following procedure:
\begin{itemize}
    \item For every value of $V_\text{AD}$, we extract the peaks in the differential conductance using the \verb|find_peaks| method from \verb|scipy| as shown in Fig.~\ref{sup:fit-method}b.
    \item We disregard the peaks that appear from the noise by identifying the most continuous lines in the signal as shown in Fig.~\ref{sup:fit-method}b,c using the code reported in the linked repository~\cite{Zenodo2025probing}.
    \item We label the extracted peaks as $\mathcal{E}_{ij}$ where the index $i$ describes the level and the index $j$ describes the voltage of the additional dot.
    \item For a given set of microscopic parameters we compute the eigenvectors and eigenvalues of each model and obtain the excitation energy spectrum $E_{ij}(\mathbf{x})$ where the indexes $i$ and $j$ are the same as before. We keep only transitions that are visible from the additional quantum dot side, that is, transitions such that $|\langle \psi_i |c_0|\psi_0\rangle| > \epsilon$ where $|\psi_0\rangle$ is the ground state, $|\psi_i\rangle$ are the excited states of the opposite parity and $\epsilon\sim10^{-3}$ is a threshold.
    \item Finally, we use an optimiser~\cite{scipy} to find the minimum of the cost function:
\end{itemize}
\begin{equation}\label{eq:loss}
    \mathcal{C} = \sum_{i,j} \left| \mathcal{E}_{ij} - E_{ij}(\mathbf{x}) \right|^2.
\end{equation}
While it is possible to fit the data for all the parameters in the model, this approach is susceptible to over-fitting given the large amount of parameters.
Therefore, we fix the chain parameters using the experimental configuration as a guide and only optimise for the unknown parameters.
Although this limits the quality of the fit, it guarantees that we do not overfit the data since the role of each interaction is well-defined.
We fix the chemical potentials $\mu_i$ and the interactions within the chain $|t_i|=|\Delta_i|$, as shown in the linked repository~\cite{Zenodo2025probing}.
For the spinless model, we fit the following microscopic parameters:
\begin{equation}
    \mathbf{x} = (U_\text{nl}, t_d, \Delta_d, \alpha_0, \mu_\text{offset}).
\end{equation}
In the case of the spinful model, we also optimise for $\theta$.
The bounds for all the optimisation parameters are sufficiently large so that we do not guide the system to a particular value.

\subsubsection{Spinless model}

\begin{figure}[h!]
    \centering
    \includegraphics[width=0.67\textwidth]{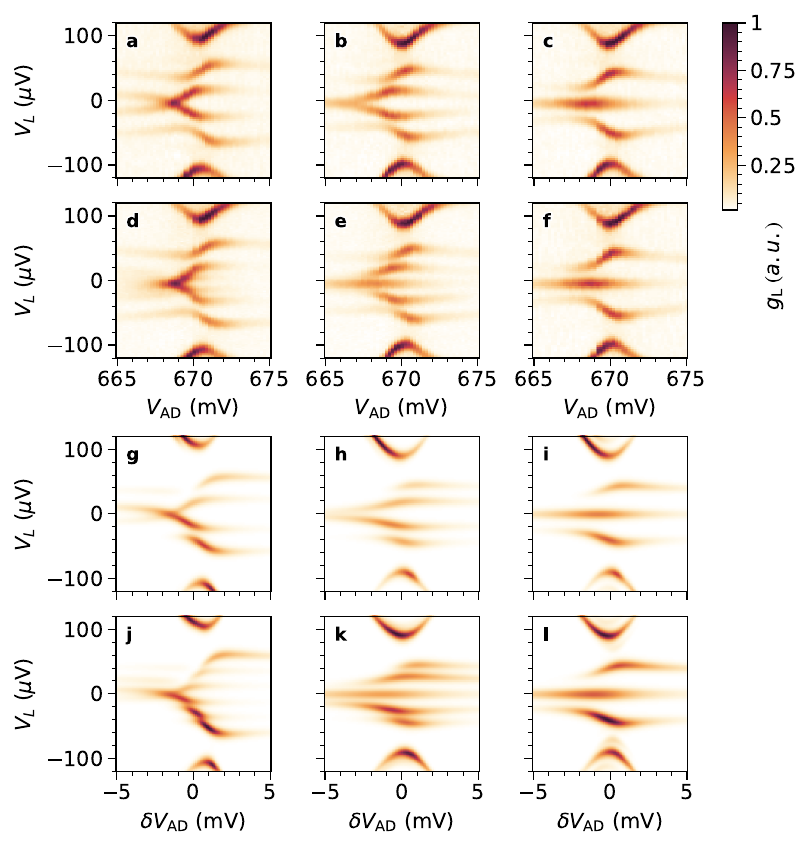}
    \caption{\textbf{Theoretical simulations of the differential conductance for the QD-test.} Transport simulations using the fitted parameters listed in Table~\ref{tab:spinless_fit}. \textbf{a-f.} Replot of the experimental data shown in Fig.~\ref{fig:4}. \textbf{g-l.} Corresponding theoretical simulations of the differential conductance, assuming a $k_B T = \SI{3}{\micro eV}$ and setting $\Gamma_R\neq\Gamma_L=\SI{5}{\micro eV}$. Both the experimental data and the simulated results were normalized and plotted in arbitrary units.}
    \label{sup:fit}
\end{figure}

We fit the datasets presented in Fig.~\ref{fig:4} using the spinless model.
The parameters extracted from the fitting are listed in Table~\ref{tab:spinless_fit} and the corresponding transport simulations are shown in Fig.~\ref{sup:fit}g-l.
We observe that the optimal solution yields systematic results for the strength of the different coupling parameters. Namely, we extract $t_d\sim \SI{50}{\micro eV}$, $\Delta_d\sim \SI{30}{\micro eV}$, and $U_\text{nl}\sim\SI{20}{\micro eV}$.

\begin{table}[h]
\centering
\begin{tabular}{|c||c|c|c|c|c|c|c||c|c|c|c|c|}
\hline
\multicolumn{1}{|c||}{ } & \multicolumn{7}{|c||}{Fixed parameters} & \multicolumn{5}{c|}{Optimisation parameters} \\
\hline
\hline
Experiment & $\mu_1$ & $\mu_2$ & $\mu_3$ & $\Delta_1$ & $t_1$ & $\Delta_2$ & $t_2$ & $U_\text{nl}$ & $t_d$ & $\Delta_d$ & $\alpha_0$ & $\mu_\text{offset}$ \\
\hline
\hline
\ref{fig:4} (a)  & 23 & 26 &  & 12 & 12 &  &  & 21 & 57 & 34 &  0.06 & -3 \\
\hline
\ref{fig:4} (b)  & 0 & 26 &  & 12 & 12 &  &  & 26 & 47 & 30 &  0.05 & 17 \\
\hline
\ref{fig:4} (c)  & 0 & 0 &  & 12 & 12 &  &  & 26 & 50 & 30 &  0.05 & 16 \\
\hline
\ref{fig:4} (e)  & 23 & 26 & 22 & 12 & 12 & 9 & 9 & 23 & 57 & 30 &  0.06 & -20 \\
\hline
\ref{fig:4} (f)  & 0 & 26 & 0 & 12 & 12 & 9 & 9 & 22 & 50 & 30 &  0.04 & 8 \\
\hline
\ref{fig:4} (g)  & 0 & 0 & 0 & 12 & 12 & 9 & 9 & 28 & 50 & 29 &  0.05 & 17 \\
\hline
\end{tabular}
\caption{\textbf{Optimised parameters for the spinless two and three-site chains.} All parameters are given in $ \SI{}{\micro eV}$ except for $\alpha_0$, which is dimensionless. The first three rows are for the two-site chain and the last three rows for the three-site chain. We optimise five parameters listed in the rightmost columns.}
\label{tab:spinless_fit}
\end{table}

\clearpage
\subsubsection{Spinful model}
In Table~\ref{tab:spinful_fit} we report the results of the spinful model corresponding to the sweet-spot cases shown in Fig.~\ref{fig:4}c,g.

\begin{table}[h]
\centering
\begin{tabular}{|c||c|c|c|c|c|c||c|c|c|c|c|c|}
\hline
\multicolumn{1}{|c||}{ } & \multicolumn{6}{|c||}{Fixed parameters} & \multicolumn{6}{c|}{Optimisation parameters} \\
\hline
\hline
Experiment & $E_Z$ & $U$ & $\eta_1$ & $\tau_1$ & $\eta_2$ & $\tau_2$ & $\theta/\pi$ &$U_\text{nl}$ & $\tau_0$ & $\eta_0$ &$\alpha_0$ & $\mu_\text{offset}$ \\
\hline
\hline
\ref{fig:4} (c)  & 200 & 3000 & 12 & 12 &  &  &  0.55 & 14 & 79 & 36 & 0.05  & 1 \\
\hline
\ref{fig:4} (g)  & 200 & 3000 & 12 & 12 & 9 & 9 &  0.51 & 10 & 75 & 39 & 0.04  & 11 \\
\hline
\end{tabular}
\caption{\textbf{Optimised parameters for the spinful chains.} All parameters are given in $ \SI{}{\micro eV}$ except for $\alpha_0$, which is dimensionless, and $\theta$, which is in units of $\pi$. We fix the interactions inside the chain and the chemical potentials at the sweet spot for a given $\theta$. We optimise six parameters listed in the rightmost columns.}
\label{tab:spinful_fit}
\end{table}
\clearpage
\begin{center}
  \textbf{\LARGE Extended Data}
\end{center}

\begin{figure}[htbp]
    \centering
    \includegraphics[width=0.7\textwidth]{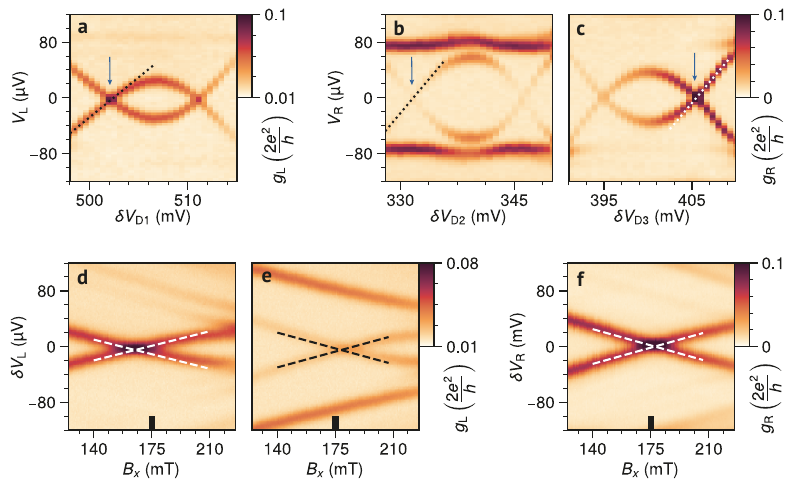}
    \caption{\textbf{QD characterization of sweet spot A. a-c.} Conductance spectroscopy as a function of $\VKi, \VKii$ and $\VKiii$. Each panel~shows a single QD orbital, measured with the other two quantum dots $\approx \SI{5}{mV}$ off-resonance. The arrows highlight the used QD resonances for sweet spot A, corresponding to a spin configuration of: $\downarrow, \downarrow, \uparrow$ respectively. From the slope of the overlaid dotted lines, we estimate the lever arms $\alpha_1, \alpha_2, \alpha_3 = 0.011, 0.013, 0.015$, respectively, where $\delta \mu_n \equiv -e \alpha_n \delta V_{\mathrm{D}n}$. $\Bx = \SI{175}{mT}$. We note the presence of high-energy states at about $\SI{80}{\micro V}$, which we attribute to the other QDs set off-resonance. \textbf{d-f} Conductance spectroscopy as a function of $B_\mathrm{x}$, the field along the nanowire. Again, each panel~is measured such that, at $\SI{175}{mT}$, only one QD is on resonance while the other two are off-resonance. From the slope of the overlaid dashed lines we estimate a $g$-factor of $\approx$ 21, 22 and 23 for the \Ki, \Kii, and \Kiii, respectively ($\EZ = g \mu_\mathrm{B} B$, where $\mu_\mathrm{B}$ is the Bohr magneton). We note that the $g$-factor can be affected by the coupling between the QDs and the neighbouring hybrids and that it might vary also as a function of the field. This can affect the estimation of the Zeeman energy. Therefore, it is important to verify that in the QD$_n$ spectra (panels \textbf{a}-\textbf{c}) there are no excited states that disperse with $\delta V_{\mathrm{D}n}$. This ensures that the Zeeman energy is larger than the scanned voltage bias range.  }
    \label{sup:QD-char-A}
\end{figure}

\begin{figure}[htbp]
    \centering
    \includegraphics[width=0.6\textwidth]{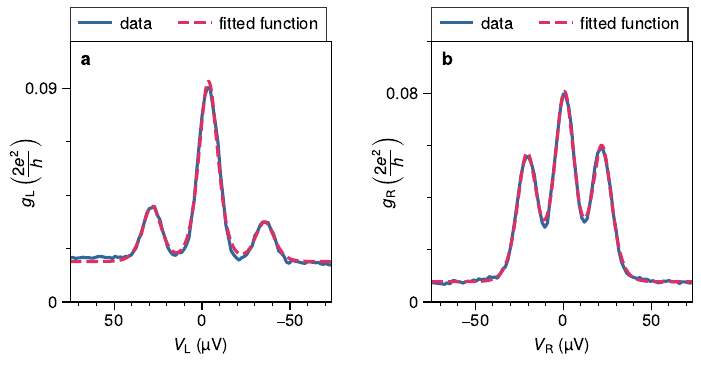}
    \caption{\textbf{Fit of the two-site chain spectra of Fig.~\ref{fig:1}f,g.} Every peak is fitted with a $\cosh$ function as suggested in Refs.~\cite{Beenakker1991theory, tenHaaf2024edge}, yielding $g_i(V_i) = A_1\cosh{\left(\frac{V_i-2t_j}{\gamma}\right)}^{-2}
+
A_2\cosh{\left(\frac{V_i}{\gamma}\right)}^{-2} 
+
A_3\cosh{\left(\frac{V_i+2t_j}{\gamma}\right)}^{-2} 
+
B_i$ where $i\in \{ \mathrm{L}, \mathrm{R} \}$ and $j\in \{ 1,2 \} $. We extract a $\abs{2t_1}=\abs{2\Delta_1}\approx \SI{32}{\micro V}$ energy gap for the left two-site chain (panel~\textbf{a}) and a $\abs{2t_2}=\abs{2\Delta_2}\approx \SI{21}{\micro V}$ energy gap for the right two-site chain (panel~\textbf{b}). Both panels yield a peak broadening $\gamma\approx \SI{7}{\micro V}$. We note that there is a finite background conductance $B_\mathrm{L} \approx 0.017 \frac{2e^2}{h}$ on the left and $B_\mathrm{R} \approx 0.008 \frac{2e^2}{h}$ on the right, which we attribute to the capacitive response of the fridge lines to the lockin excitations \cite{Wang2024Spatial}.}
    \label{sup:fit-gap}
\end{figure}

\begin{figure}[htbp]
    \centering
    \includegraphics[width=\textwidth]{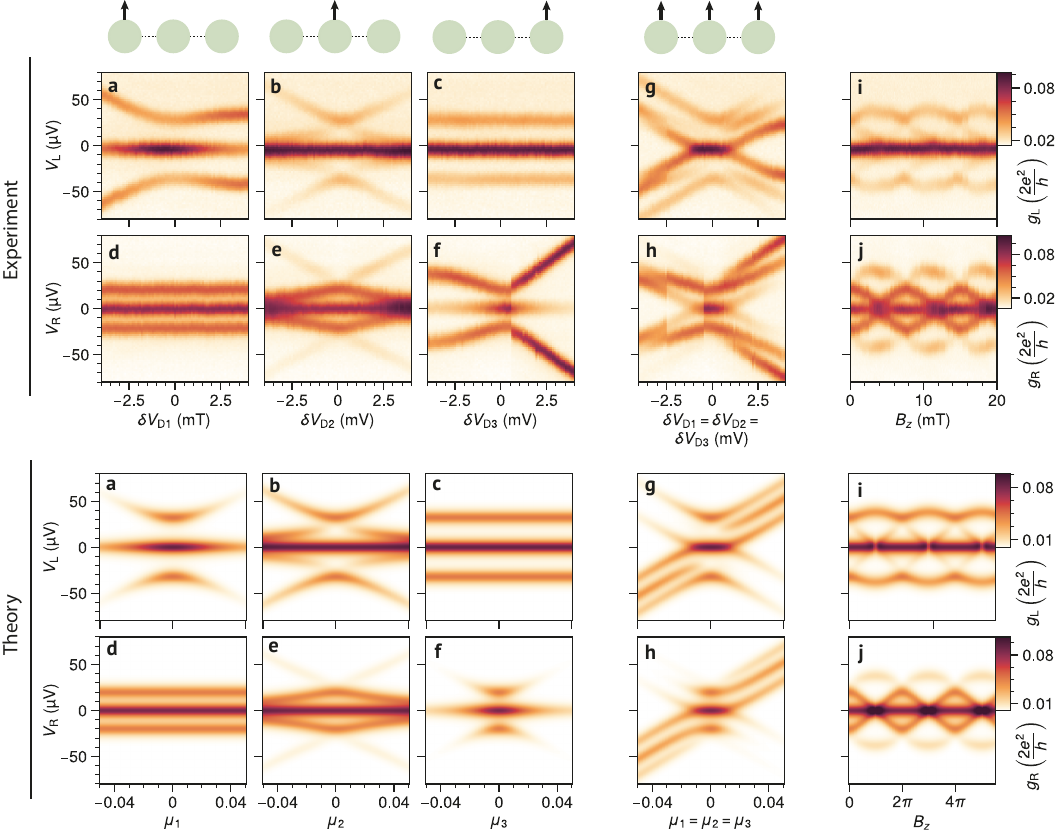}
    \caption{\textbf{Experiment.} Conductance spectroscopy of the three-site chain presented in Figs.~\ref{fig:2}e and~\ref{fig:3}. \textbf{a-f.} The spectrum, measured from both the left and right lead, as a function of each QD making up the chain. In all panels, only a single excited state is observed unless either the \Kii or the phase is detuned. \textbf{g, h.} The spectrum measured from the left and right lead, as a function of all QDs of the chain detuned simultaneously. Two excited states are visible when all QDs are detuned, yet only one excited state remains when all QDs are aligned at zero energy. \textbf{i, j.} The spectrum measured from the left and right lead, as a function of the out-of-plane field $B_\mathrm{z}$. \textbf{Theory.} \textbf{a-j.} Corresponding theoretical simulations of the differential conductance of a three-site chain. Calculations are performed following the scattering matrix approach \cite{Dvir2023realization}, assuming a finite temperature of $k_\mathrm{B}T=\SI{3}{\micro eV}$ and a finite coupling to the normal leads: $\Gamma_\mathrm{L} = \Gamma_\mathrm{R} = \SI{0.7}{\micro eV}$, while using the extracted coupling amplitudes of Fig.~\ref{sup:fit-gap}; $\abs{t_\mathrm{1}} = \abs{\Delta_\mathrm{1}} = \SI{16}{\micro eV}$ and $\abs{t_\mathrm{2}} = \abs{\Delta_\mathrm{2}} = \SI{10}{\micro eV}$.}
    \label{sup:20-pack-A}
\end{figure}

\begin{figure}[htbp]
    \centering
    \includegraphics[width=0.75\textwidth]{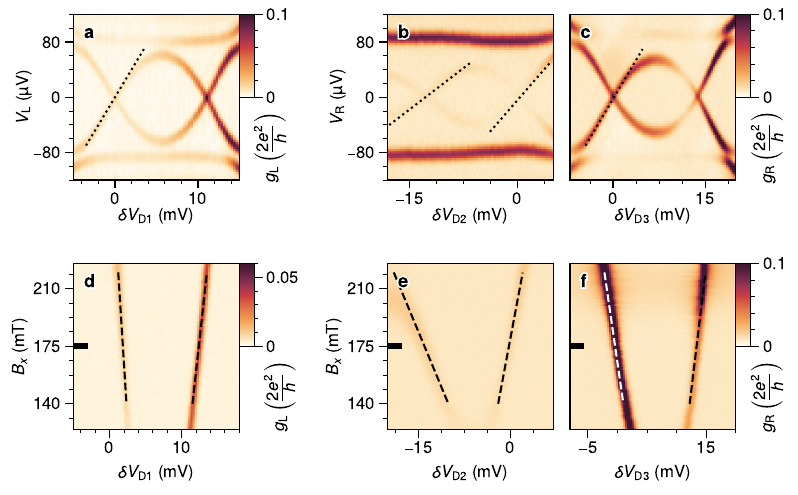}
    \caption{\textbf{QD characterization of sweet spot B. a-c.} Conductance spectroscopy as a function of $\VKi, \VKii$, and $\VKiii$. Each panel~shows a single QD orbital, measured at $B_\mathrm{x}=\SI{175}{mT}$. $\delta V_{\mathrm{D}n} = 0$ denotes the QD resonance making up the Kitaev chain for sweet spot B, corresponding to a spin configuration of; $\downarrow, \uparrow, \downarrow$. The lever arm of each QD plunger gate is estimated from the slope of the overlaid dotted lines, corresponding to $\alpha_1, \alpha_{2\downarrow}, \alpha_{2\uparrow}, \alpha_3 = 0.020, 0.08, 0.012, 0.015$ for $\Ki$, the left resonance of $\Kii$, the right resonance of $\Kii$,
    and $\Kiii$, respectively. \textbf{d-f.} Zero-bias conductance as a function of $B_\mathrm{x}$, the field along the nanowire, and $\VKi, \VKii, \VKiii$, respectively. From the slope of the overlaid dashed lines we estimate a $g$-factor of $\approx$ 15, 25 and 20 for \Ki, \Kii, and \Kiii, respectively.}
    \label{sup:QD-char-B}
\end{figure}

\begin{figure}[htbp]
    \centering
    \includegraphics[width=\textwidth]{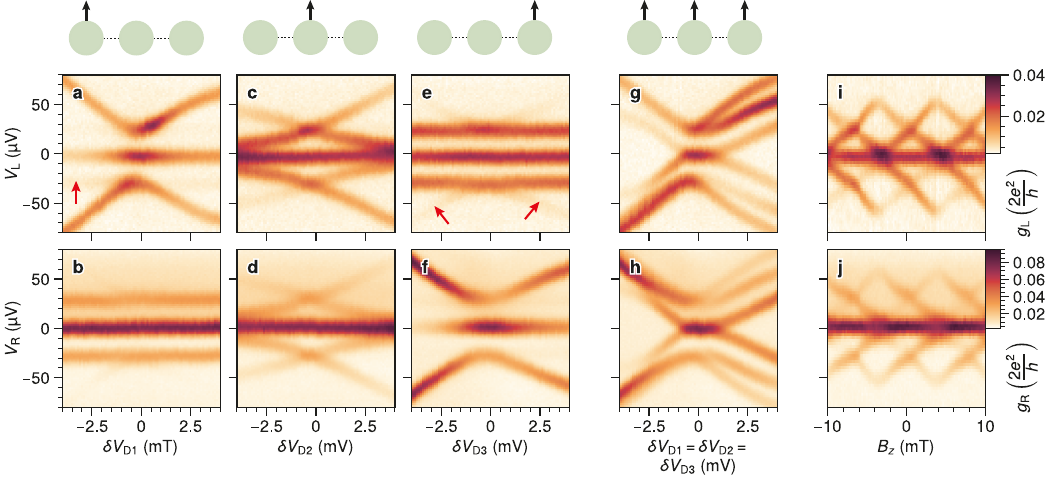}
    \caption{\textbf{Characterization of the three-site chain spectra in sweet spot B, reproduction of Fig.~\ref{fig:3}. a-f.} The spectrum, measured from the left and right lead, as a function of each QD making up the chain. We note that in some panels a second excited state is faintly visible (see red arrows in panels \textbf{a} and \textbf{e}). This could be due to the lower Zeeman energy (compare Fig. \ref{sup:QD-char-B} and \ref{sup:QD-char-A}) possibly enabling next-nearest-neighbour coupling or to a less precise identification of the field corresponding to $\varphi = 0$, here set at $B_\mathrm{z} = \SI{0.8}{mT}$.  
    \textbf{g, h.} The spectrum measured from the left and right lead, as a function of all QDs of the chain detuned simultaneously. \textbf{i, j.} The spectrum measured from the left and right lead, as a function of the out-of-plane field $B_\mathrm{z}$.}
    \label{sup:10-pack-B}
\end{figure}

\begin{figure}[htbp]
    \centering
    \includegraphics[width=\textwidth]{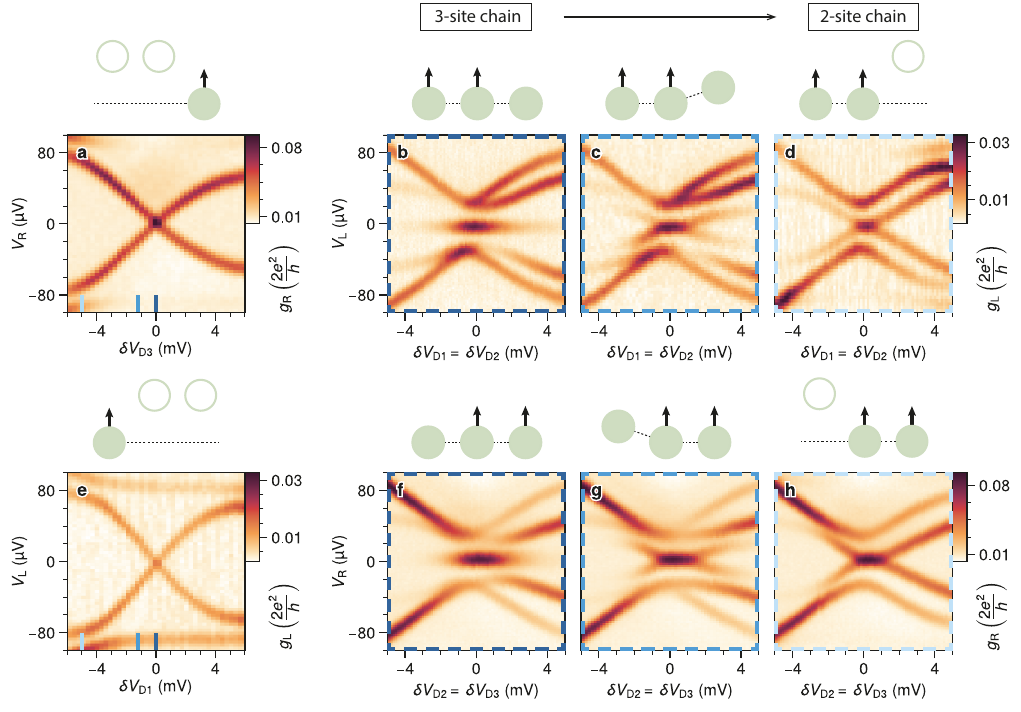}
    \caption{\textbf{Transition from a three-site chain to a two-site chain as one quantum dot is detuned.} Here we show the measured conductance spectra for sweet spot B for different detunings of the outer QD, demonstrating the ability to tune from the three-site to the two-site regime by moving one of the outer QDs off-resonance. \textbf{a.} Conductance spectroscopy $\gR$, measured from the right lead as a function of $\VKiii$, revealing the sub-gap density of states of \Kiii. The panel~is centered around a single charge degeneracy point. \Ki and \Kii were placed off-resonance during this measurement. Blue insets denote the position of $\VKiii$ in subsequent panels \textbf{b}, \textbf{c} and \textbf{d}. \textbf{b-d.} The spectrum of the chain measured from the left lead as a function of $\VKi = \VKii$ at three different positions of $\VKiii$. In panel~\textbf{b}, \Kiii is positioned on resonance, restoring the three-site chain when $\delta \VKi = \delta \VKii = 0$. This is evident from the persistent zero-bias peak upon detuning $\delta \VKi$ and $\delta \VKii$. In panel~\textbf{c}, $\VKiii$ is slightly detuned. The zero-bias peak is now split when $\delta \VKi$ and $\delta \VKii$ are detuned. Finally, in panel~\textbf{d}, $\VKiii$ is placed off-resonance. Judging from panel~\textbf{a}, its chemical potential is now $\approx \SI{80}{\micro V}$ ($>\abs{t_2}=\abs{\Delta_2}$) such that it no longer plays a significant role in the low-energy spectrum of the chain. Therefore, we argue that the chain at that point can be approximated as a two-site chain. \textbf{e-h.} The same experiment performed from the opposite side of the chain. When placing $\VKi$ far off-resonance (panel~\textbf{h}), the right side of the device converts to a two-site chain.}
    \label{sup:3-2-transition}
\end{figure}

\begin{figure}[htbp]
    \centering
    \includegraphics[width=\columnwidth]{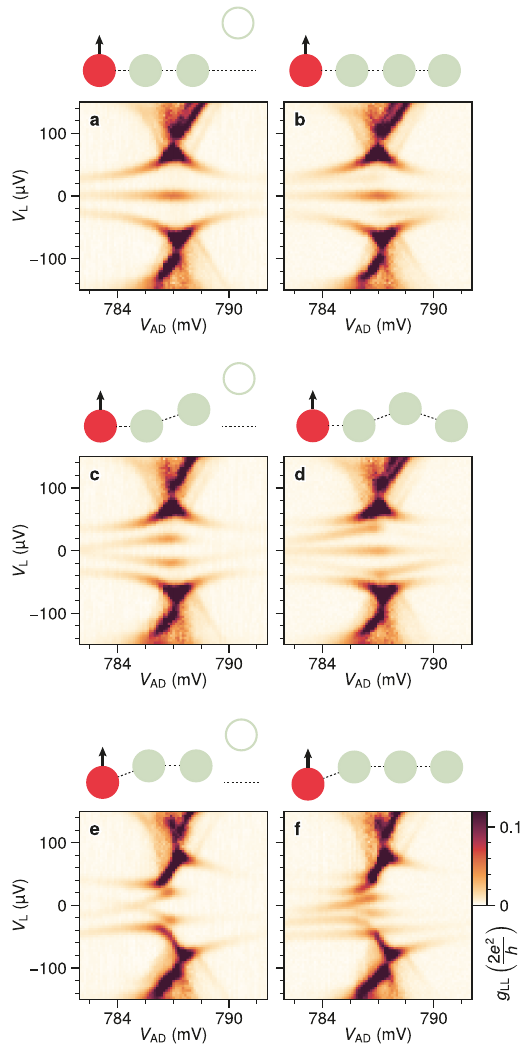}
    \caption{\textbf{QD-test for sweet spot B, reproduction of Fig. ~\ref{fig:4}. a-b.} QD-test at the sweet spot for the two-site and three-site chain, respectively, similar to Fig. \ref{fig:4}c,g. \textbf{c-f.} QD-test for the detuned two-site and three-site chain, resembling Fig. \ref{fig:4}b,f,a,e, respectively.}
    \label{sup:PC-spin2}
\end{figure}

\begin{figure}[htbp]
    \centering
    \includegraphics[width=\columnwidth]{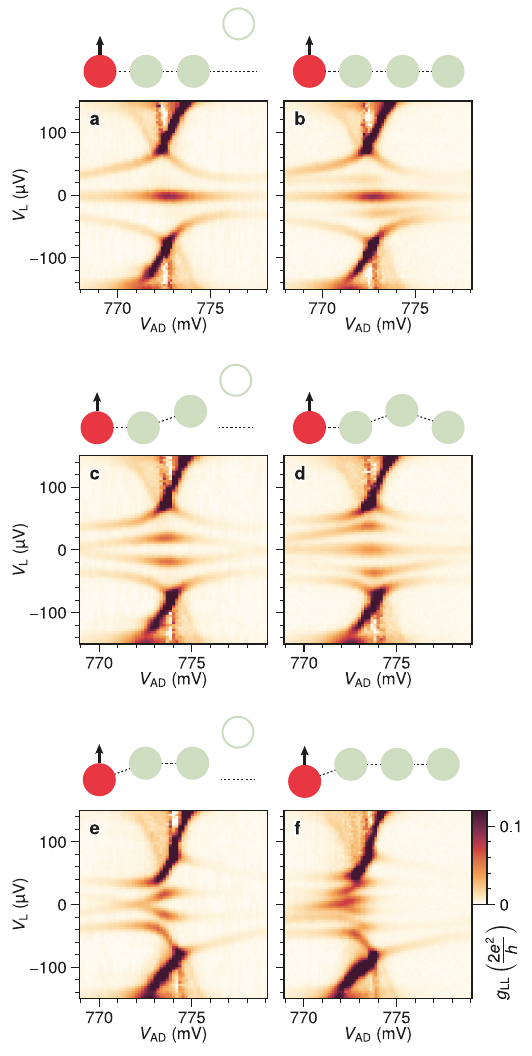}
    \caption{\textbf{QD-test for sweet spot B with altered spin configuration.} Same as Fig.~\ref{sup:PC-spin2} but using a different spin in the additional quantum dot.}
    \label{sup:PC-spin1}
\end{figure}

\begin{figure}[htbp]
    \centering
    \includegraphics[width=0.95\textwidth]{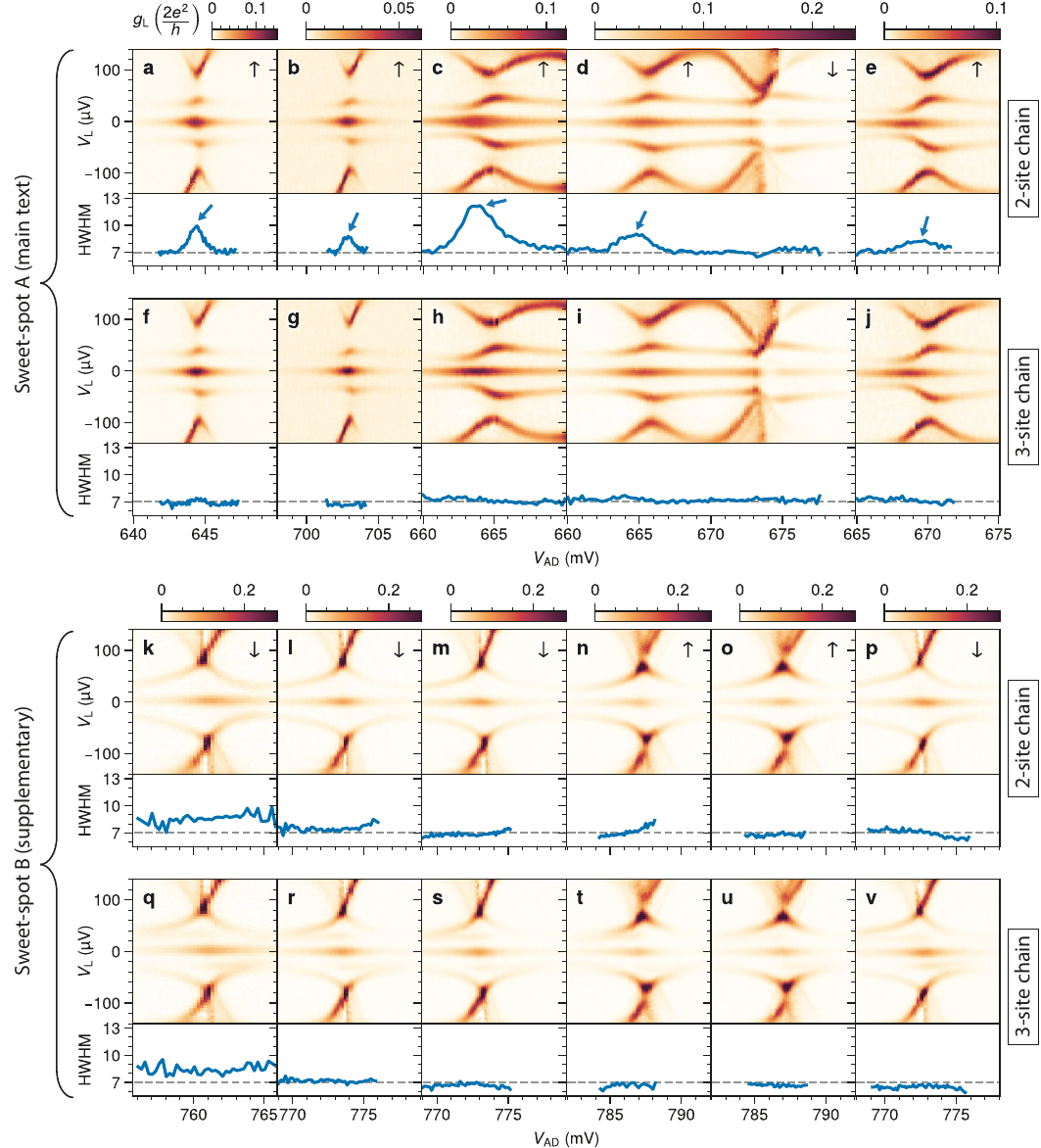}
\caption{\textbf{Inventory of all the QD-tests comparing two- and three-site chains.} Here we report all measured QD-tests comparing two- and three-site chains at the sweet-spot. The first two rows are measured on sweet spot A (studied in Figs.~\ref{fig:1} to~\ref{fig:4} and \ref{sup:QD-char-A} to~\ref{sup:20-pack-A}) whilst the last two rows are measured on the sweet spot B (Figs.~\ref{sup:QD-char-B} to~\ref{sup:PC-spin1}). Even rows study three-site chains while odd rows study the corresponding two-site chains after setting \Kiii off-resonance. For each sweet spot, multiple iterations of the QD-test were performed. Each iteration corresponds to a different tune-up of the additional quantum dot or to a fine adjustment in the sweet-spot centering.
    Panels \textbf{e} and~\textbf{j} report the same data of Fig.~\ref{fig:4}c,g; 
    panels \textbf{o} and~\textbf{u} report the same data of Fig.~\ref{sup:PC-spin2}a,b;
    panels \textbf{p} and~\textbf{v} report the same data of Fig.~\ref{sup:PC-spin1}a,b.
    For every QD-test, the black arrow in the top-right corner indicates the electron spin in the additional QD, it is inferred from the comparison with the theory model.\newline
    Below every conductance spectrum, we plot the half-width at half-maximum (HWHM) of the zero-bias peak (ZBP), wherever the ZBP height is at least $0.01 \frac{2e^2}{h}$ (otherwise the signal is too low to extract the HWHM reliably). Whenever the additional quantum dot is off-resonance, we measure $\mathrm{HWHM} = 7 \pm \SI{1}{\micro V}$, apart from panels \textbf{k} and \textbf{q} where it is slightly larger due to lower $\VL$ resolution (see code in the linked repository~\cite{Zenodo2025probing}). Conversely, when AD is brought into resonance, we sometimes resolve a different behaviour for two- and three-site chains: the latter always show $\mathrm{HWHM} = 7 \pm \SI{1}{\micro V}$, whereas for two-site chains there is sometimes an excess in the measured HWHM, highlighted by the blue arrows. We attribute this excess in the measured HWHM to an imperfect centering at the sweet spot.}
    \label{sup:HWHM}
\end{figure}

\begin{figure}[htbp]
    \centering
    \includegraphics[width=0.6\textwidth]{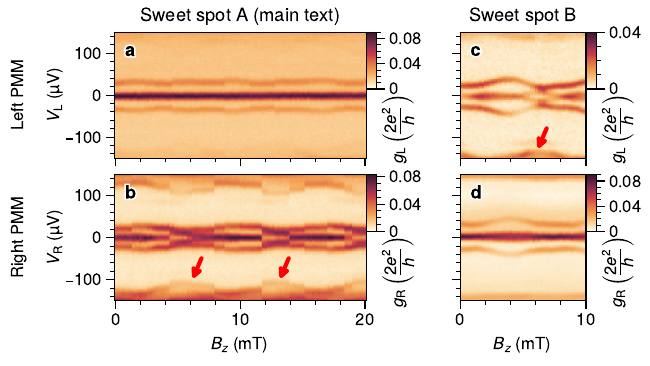}
    \caption{\textbf{Phase dependence of two-site Kitaev chains.} The left column (panels \textbf{a} and~\textbf{b}) concerns sweet spot A (discussed in Figs.~\ref{fig:1} to~\ref{fig:4} and \ref{sup:QD-char-A} to~\ref{sup:20-pack-A}); the second column (panels \textbf{c} and~\textbf{d}) concerns sweet spot B (characterized in Figs.~\ref{sup:QD-char-B} to~\ref{sup:PC-spin1}). In the first row, \Kiii is set off-resonance to define a two-site chain on the left side of the device. In the second row, \Ki is set off-resonance (with \Kiii back on resonance), to define a two-site chain on the right side.
    We note that the measured two-site conductance spectra show some phase dependence. This could be attributed to the corresponding modulation of the energy of the Andreev bound states (ABSs) populating the hybrid regions. They are marked with red arrows in panel \textbf{b} and \textbf{c}. Since the ABSs influence the $t_n$ and $\Delta_n$ couplings~\cite{Liu2022tunable, Bordin2023tunable, TorresLuna2024flux, Kulesh2025A}, their modulation can affect the $\abs{t_n}=\abs{\Delta_n}$ sweet-spot condition (see for instance panel \textbf{c}, where the ZBP is split at $\approx \SI{6}{mT}$). 
    Hence, the sweet-spot condition of three-site chains might be slightly imprecise if $\Bz$ is varied (Figs.~\ref{fig:2}e,f,g and~\ref{sup:0-π-inventory}). When $\Bz$ is fixed (all other figures), we optimise for an accurate $\abs{t_n}=\abs{\Delta_n} \; \forall n$.}
    \label{sup:PMM-phase-dependence}
\end{figure}

\begin{figure}[htbp]
    \centering
    \includegraphics[width=\textwidth]{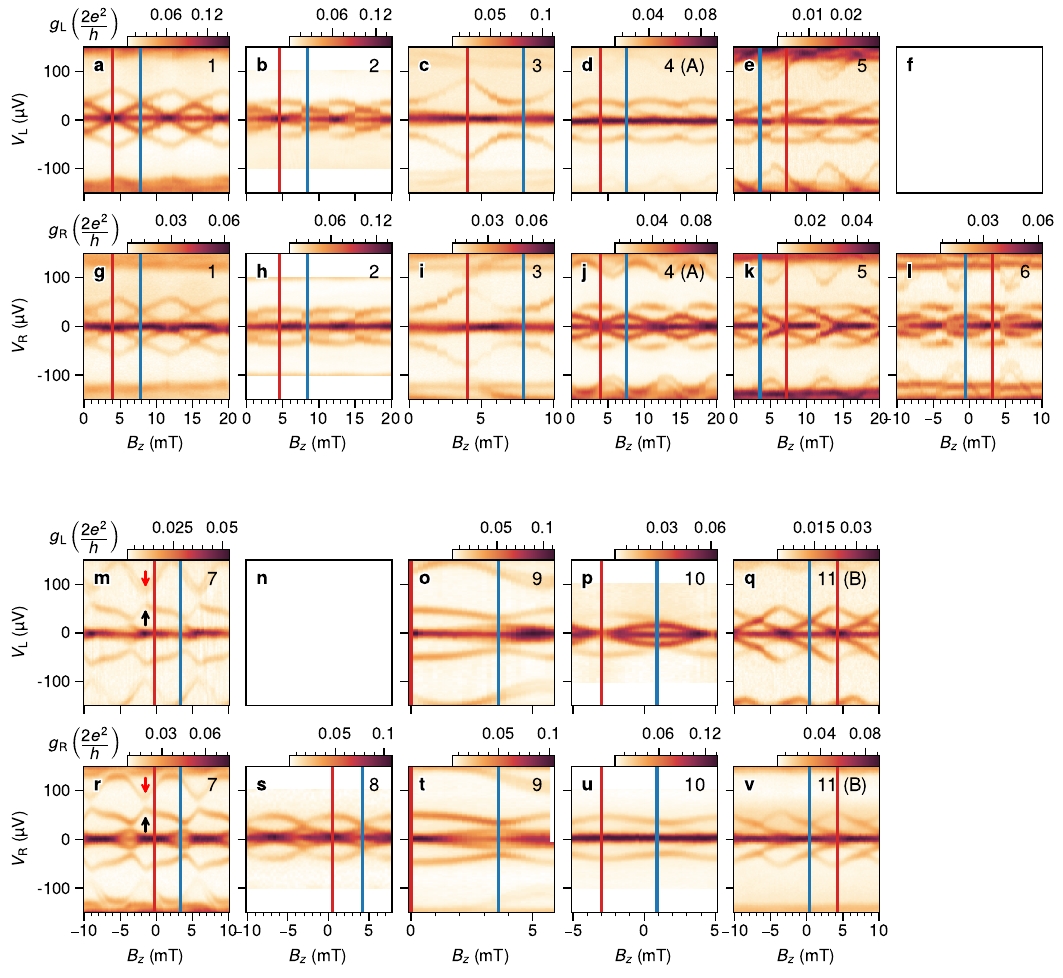}
    \caption{\textbf{Full catalog of three-site Kitaev chain spectra as a function of \textbf{\textit{B}}$_z$.} Overall, we measured the $\Bz$ dependence of 11 different three-site sweet spots; they are labelled with the numbers in the top left corner of each panel. Different sweet spots correspond to different tune-ups of the device, hence, they can differ in both the QD orbitals and the hybrid settings. Different sweet spots were characterized to different extents, for instance, we performed QD-tests only on sweet spots 4 (i.e. ``A'') and 11 (i.e. ``B''). Panels~\textbf{f} and~\textbf{n} are white because we did not measure the left conductance in those cases. In all other panels, blue lines identify the field where $\varphi=0$, red lines identify the field where $\varphi = \pi$. The 0 and $\pi$ points are identified manually by comparing the measured spectra with simulations. The identification was repeated, independently, by two operators, yielding an average difference of $\SI{0.2}{mT}$ (see code in the linked repository~\cite{Zenodo2025probing}). We note that some of the spectra appear distorted. For instance, in panels~\textbf{m} and~\textbf{r} the second excited state appears anomalously high in energy (black arrow) near the minimum ABS energy (red arrows). This is consistent with what is discussed in Fig.~\ref{sup:PMM-phase-dependence}: a lower ABS energy can increase the $t_n$ and $\Delta_n$ amplitudes~\cite{Bordin2023tunable}. Finally, we note that the period is $\approx \SI{7.5}{mT}$ for all 11 cases, corresponding to a loop area of $\approx\SI{0.28}{\micro m^2}$. However, the internal area of our superconducting loop is $\approx \SI{0.13}{\micro m^2}$. The discrepancy suggests that the field penetrates the Al film.}
    \label{sup:0-π-inventory}
\end{figure}

\bigskip
\begin{itemize}
    \item Figs.~\ref{sup:QD-char-A} to~\ref{sup:20-pack-A} concern the ``sweet spot A'' presented in the main text.
    \item Figs.~\ref{sup:QD-char-B} to~\ref{sup:PC-spin1} report the reproduction of the main results on a another sweet spot (``sweet spot B'').
    \item Figs.~\ref{sup:HWHM} and~\ref{sup:PMM-phase-dependence} compare sweet spots A and B.
    \item Fig.~\ref{sup:0-π-inventory} reports all the measured phase dependences of 11 different sweet spots we found in the reported device. Different sweet spots differ in either hybrid or QD gate voltages, or both, and were measured to very different extents. QD-tests were performed only for the fourth and the eleventh sweet spot, corresponding to sweet spots A and B, respectively.
\end{itemize}
\bigskip

\clearpage
\enlargethispage{2cm}

\clearpage

\clearpage
\twocolumngrid

\clearpage
\onecolumngrid


\end{document}